\documentclass[12pt]{article}
\usepackage{multirow}
\usepackage{booktabs}
\usepackage{dsfont}
\usepackage{subcaption}
\usepackage{amssymb}
\usepackage{amsmath}
\usepackage{amstext}
\usepackage{amsthm}
\usepackage{graphicx} 
\usepackage{color,verbatim}
\usepackage{mathtools}
\usepackage{enumitem}
\usepackage{algorithm}
\usepackage{algpseudocode}
\usepackage{chngcntr}
\usepackage{xr}
\usepackage[bookmarks=false,
colorlinks,
linkcolor=blue,
anchorcolor=blue,
citecolor=blue,
filecolor=blue,
urlcolor=black
]{hyperref}

\selectfont

\DeclareMathOperator{\sign}{sign}
\DeclareMathOperator{\argmax}{argmax}

\def\squarebox#1{\hbox to #1{\hfill\vbox to #1{\vfill}}}
\def\boxit#1{\vbox{\hrule\hbox{\vrule\kern6pt
			\vbox{\kern6pt#1\kern6pt}\kern6pt\vrule}\hrule}}

\def\sumk0p{\sum_{k=0}^\ell}

\makeatletter
\newcommand{\namedlabel}[2]{%
	\begingroup
	#2%
	\def\@currentlabel{#2}%
	\def\@currentlabelname{#2}%
	\phantomsection
	\label{#1}%
	\endgroup
}
\makeatother

\renewcommand{\tilde}{\widetilde}
\renewcommand{\hat}{\widehat}

\graphicspath{{./Figures/}}
\numberwithin{equation}{section}
\allowdisplaybreaks

\newtheoremstyle{break}
  {9pt}
  {12pt}
  {\upshape}
  {}
  {\bfseries}
  {.}
  {\newline}
  {}

\newtheorem{theo}{Theorem}

\newtheorem{defi}{Definition}

\theoremstyle{definition}

\theoremstyle{break}

\theoremstyle{remark}

\usepackage{natbib}
\title{•}
\author{}
\date{}
\begin{document}
\begin{center}
{\bf \Large 
Semi-supervised Classification for Noisy Functional Data with Application to Astronomical Spectra} \\[1cm]
\end{center}

\begin{center}
Ruoxu Tan \\
School of Mathematical Sciences and School of Economics and Management, Tongji University,
ruoxut@tongji.edu.cn\\ 
Mingjie Jian\\
Institute of Astronomy, University of Cambridge, jian-mingjie@outlook.com\\
Yiming Zang \\
Department of Sciences,  
North China University of Technology,
yiming.zang@ncut.edu.cn\\
\end{center}

\linespread{1}\selectfont 
\begin{abstract}
Despite its extensive development for multivariate data, semi-supervised learning remains underdeveloped for functional data, especially under discrete and noisy observations. We develop a
density-sensitive semi-supervised framework for functional data
supported on a low-dimensional manifold by adapting the
Fermat distance to reconstructed trajectories. The
resulting pairwise distances are used to construct a weighted $k$-nearest-neighbor classifier and multidimensional-scaling-based classifiers.  To accommodate massive datasets commonly seen in semi-supervised applications, we design a computationally efficient estimation procedure tailored for discrete and noisy functional observations. Theoretically, we establish exponentially decaying convergence rates of the $k$-NN classifier and the consistency of the estimated Fermat distance. Crucially, our results reveal that incorporating unlabeled data may not lead to improved classification accuracy without a sufficiently fast-growing individual sampling rate, precisely due to discrete and noisy observations. In most simulation settings satisfying the manifold and cluster assumptions, the proposed classifiers outperform the supervised benchmarks considered; in the Gaia spectra analysis, they attain higher agreement with high-confidence proxy labels.
\end{abstract}

\linespread{1.7}\selectfont
\section{Introduction}\label{sc_intro}
Functional data arise when each observational unit is represented by a curve, spectrum, or time-indexed process rather than a finite-dimensional vector. Such data are now routine in astronomy \citep{RecioBlanco2023}, chemometrics \citep{Wohlers2026}, and neuroscience \citep{Poldrack2024}, demanding advanced development in functional data analysis \citep{Wang2016}.
Similar to the analysis of standard data objects, machine learning tasks, such as classification and clustering, are important components in functional data analysis. In the task of classification, the aim is to train a classifier based on a dataset consisting of paired functional observations and their classes. This supervised learning paradigm has been studied extensively in the literature \citep{Delaigle2012,Dai2017,Berrendero2018,Tan2025}; see \citep{Wang2024} for a review.  
 
In contrast, semi-supervised learning, in which only a portion of individuals are labeled, remains comparatively underdeveloped in functional data analysis, despite its extensive development for high-dimensional data \citep{Cai2020,VanEngelen2020,Chakrabortty2022,Deng2024}. Specifically, existing work on semi-supervised learning for functional data is largely application-specific
\citep{Marchal2025,Wohlers2026}, while general-purpose
semi-supervised approaches, such as modified support vector
machines \citep{Ding2017}, do not explicitly account for measurement error \citep{Rossi2006}. 

Yet, functional data applications under the semi-supervised paradigm indeed occur in many scientific fields. Our motivating application comes from Gaia, an ambitious space observatory launched by the European Space Agency to map the Milky Way \citep{GaiaCollaboration2016,GaiaCollaboration2023}. Typically, the observations of each star include photometric time series, radial velocity spectrometer mean spectra, etc., based on which astronomers classify them into a few classes such as solar-like and binary stars. However, the number of observations in Gaia is astronomical (more than one billion), and thus it is impossible for all stars to be labeled by experts. A standard remedy is to train a machine learning algorithm based on a small set of expert-labeled data to predict the rest. For example, \citet{Rimoldini2023} classified 12.4 million sources into 25 variability classes using a training sample cross-matched with
existing astronomical catalogs. Each predicted class is
accompanied by a confidence score. Focusing on a subset from \citet{Rimoldini2023}'s results by a bound on the signal-to-noise ratio and fewer classes, we treat the radial velocity spectrometer mean spectra with the score larger than a threshold as labeled data and the rest as unlabeled data. The key question arises: Can the unlabeled sample improve classification accuracy if it is properly utilized in the training stage? 

To answer this question, we build our methodology on two structural hypotheses: the manifold and cluster assumptions. The former assumes that functional data lie on an unknown low-dimensional Riemannian manifold \citep{Chen2012,Lin2021,Tan2024}, and the latter requires data lying in the same high-density region tend to possess the same label, which is commonly assumed in the literature of semi-supervised learning \citep{Rigollet2007,Azizyan2013}. These two assumptions are plausible for astronomical spectra due to their shape features. Standard Euclidean or geodesic distances do not account for data density, thus may be less effective under the cluster assumption. In contrast, the cornerstone of our methodology is the functional version of the Fermat distance, which is a density-sensitive metric aligning with the cluster assumption \citep{Hwang2016,Little2022,Groisman2022,Fernandez2023,Trillos2024}. Specifically, our main contributions can be summarized as follows: 
\begin{itemize}
    \item We extend the Fermat distance to the context of functional data. Based on this, we propose novel semi-supervised functional classifiers including the weighted $k$-nearest neighbors (NN) classifier and multidimensional scaling (MDS)-induced classifiers.

    \item We develop a detailed estimation procedure designed for discrete and noisy functional observations. Our procedure is computationally efficient, which is crucial for modern semi-supervised applications where the data are usually massive.

    \item In theory, we establish the consistency of the estimated Fermat distance under noisy and discrete functional observations, and derive exponentially decaying convergence rates of the expected excess risk within clusters of the weighted $k$-NN classifiers, using either true or estimated Fermat distances. In particular, our theoretical results quantify the effect of trajectory reconstruction, which indicates that the unlabeled sample may not be beneficial without a sufficiently fast-growing individual sampling rate.
    
\end{itemize}
Our theoretical finding on estimated Fermat distances exposes unique features of error-contaminated functional data, as it has no counterpart for error-free high-dimensional data. The key dilemma here is that the estimation error of the Fermat distance is aggregated with larger sample sizes due to recovering functional trajectories, thus demanding more accurate reconstruction of each trajectory. We provide sufficient conditions for the reconstruction term to be asymptotically negligible. Although the theoretical results based on upper bounds are not sufficient to prove the unlabeled sample is useless under the sparse or moderate design, we empirically illustrate by simulated examples that the unlabeled sample improves classification accuracy only when the individual sampling rate is relatively large. In addition, we also conduct extensive experiments including: (1) demonstrate via simulated examples and the astronomical dataset that our semi-supervised classifiers outperform existing supervised benchmarks; and (2) test the computational efficiency and robustness of the tuning parameters.
  
The rest of the article is organized as follows. In Section~\ref{sc_MM}, we introduce the model and data, followed by the Fermat distance for functional data and our classification strategies. Section~\ref{sc_est} details our estimation procedure based on noisy and discrete functional observations. The theoretical consistency and convergence results are provided in Section~\ref{sc_theo}. The experiments on simulated data and the astronomical data are provided in Sections~\ref{sc_sim} and \ref{sc_real}, respectively. We conclude with a discussion in Section~\ref{sc_dis}. The technical proofs and additional experimental results are provided in the Supplementary Material.

\section{Model and Methodology}\label{sc_MM}

\subsection{Model and Data}\label{sc_model}
Let $X=X(\cdot)$ denote a second-order random function on a compact interval, which we rescale to $[0,1]$ without loss of generality. Let $Y\in\mathcal{Y}=\{0,\ldots,M-1\}$ denote the label variable of $M$ classes. The set $\{(X_i,Y_i)\}_{i=1}^{n_\ell}$ contains independent and identically distributed (iid) copies of $(X,Y)$. In practice, instead of observing $X_i$ directly, we often only observe a discrete and noisy version $(T_{ij},\tilde{X}_{ij})$, for $j=1,\ldots,J_i$, satisfying
\begin{align}\label{eq_discrete}
\tilde{X}_{ij} = X_i(T_{ij})+\epsilon_{ij}\,,
\end{align}
where $T_{ij}\in[0,1]$ denotes the $j$-th observation time and
$\epsilon_{ij}$ is a mean-zero measurement error with $E(\epsilon_{ij}^2)< \infty$. We assume that
the errors $\epsilon_{ij}$ are mutually independent and independent of $(X_i,Y_i)$. The design points $T_{ij}$ may be fixed or random; in the
latter case, their density is assumed to be bounded away from zero on $[0,1]$. In our application of the spectral data, the $T_{ij}$'s are fixed, uniform, and sufficiently dense; see Section~\ref{sc_real_des} for details. It follows that the labeled sample consists of $\{(T_{ij},\tilde{X}_{ij})_{j=1}^{J_i},Y_i\}_{i=1}^{n_\ell}$. 

In addition to the labeled sample, we also observe an abundant unlabeled sample $\{(T_{ij},\tilde{X}_{ij})_{j=1}^{J_i}\}_{i=n_\ell+1}^{n}$, which is a discrete and noisy version of $\{X_i\}_{i=n_\ell+1}^{n}$ satisfying model~\eqref{eq_discrete} with $n=n_\ell+n_u$. Here, $X_i$ from the unlabeled sample is assumed to follow the same distribution as $X$. The unlabeled sample size $n_u$ is often much larger than the labeled sample size $n_\ell$. This is particularly the case in modern astronomical applications, where vast quantities of astronomical observations are collected automatically, but expert annotation remains expensive and limited. 

To effectively exploit both the labeled and unlabeled samples for classification, we rely on two structural hypotheses: the manifold and cluster assumptions. First, to characterize the intrinsic structure of $X$ that is plausibly beneficial for classification, we assume that $X$ lies on an unknown compact, connected $d$-dimensional Riemannian manifold $\mathcal{M}$ without boundary that is smoothly embedded in the ambient Hilbert space $\mathcal{L}^2([0,1])$. Consequently, the geodesic distance $d_{\mathcal{M}}$ is induced from the inner product in $\mathcal{L}^2([0,1])$. The manifold assumption implies that the data exhibit nonlinearity and low dimensionality, which are often accompanied with visual characteristics. For example, phase variation, i.e., horizontal perturbations of prominent shape features, is a common source of nonlinearity \citep{Chen2012,Tan2024}; in addition, sufficient smoothness often corresponds to a low-dimensional structure. These characteristics are commonly seen in functional data applications \citep{Wang2016,Wang2024}. For astronomical spectra, phase variations and fluctuations in absorption lines render the manifold assumption plausible. In contrast, functional trajectories lacking shared shape features and dominated by high frequencies and  uncorrelated variations would be more suitably modeled by standard Gaussian processes.

Second, incorporating the unlabeled sample can be beneficial for classification only if the marginal distribution $P_X$ is informative about the conditional distribution $P_{Y|X}$. We formalize this via the cluster assumption:  If two points $X_1$ and $X_2$ reside in the same high density region, i.e., a cluster, their labels $Y_1$ and $Y_2$ are more likely to be identical. This assumption will be stated formally in Section~\ref{sc_theo}. There, we show that the individual sampling rate $J_i$ is required to tend to infinity sufficiently fast so that the graph-aggregated reconstruction error to be asymptotically negligible. Therefore, we focus on the dense design, i.e., $J_i\to\infty$ as $n\to\infty$ for all $i$.

Our goal is to learn a robust and accurate classifier $g:\mathcal{L}^2([0,1])\to \mathcal{Y}$ by leveraging both the labeled sample $\{(T_{ij},\tilde{X}_{ij})_{j=1}^{J_i},Y_i\}_{i=1}^{n_\ell}$ and the unlabeled sample $\{(T_{ij},\tilde{X}_{ij})_{j=1}^{J_i}\}_{i=n_\ell+1}^{n}$.

\subsection{Fermat Distance}
The classical geodesic distance $d_\mathcal{M}$ has been used extensively in the literature of manifold learning since the seminal work of \citet{Tenenbaum2000}. However, $d_\mathcal{M}$ only concerns the intrinsic geometric structure of the support of the covariate $X$, and crucially, it does not take into account the density of $X$. Recall from the cluster assumption that data residing in a high density region are more likely to possess the same label. It turns out that the geodesic distance does not exploit the cluster assumption, and thus it may be less effective for semi-supervised classification.

Consequently, a density-sensitive metric is desired under the context of semi-supervised classification. In the case of manifold-support data, the density-sensitive metric that is properly aligned with the cluster assumption is the so-called Fermat distance \citep{Groisman2022,Fernandez2023,Trillos2024}. Next, we extend the Fermat distance originally restricted to random vectors to random functions. 

Let $\|\cdot\|_{\mathcal{L}^2}$ denote the $\mathcal{L}^2$ distance in $\mathcal{L}^2([0,1])$ and $f_X$ denote the density of $X$. Recall that $X$ is assumed to lie on a finite-dimensional compact and connected Riemannian manifold $\mathcal{M}$ without boundary, then it is reasonable to assume that the density $f_X$ exists with respect to the volume measure on $\mathcal{M}$ \citep{Lin2021}. Now for any $\alpha\geq 1$ and two points $x_1,x_2$ on $\mathcal{M}$, we define the power-$\alpha$ Fermat distance as
\begin{align}\label{eq_fd}
d_{\mathcal{M},\alpha}(x_1,x_2) =  \mu_{\alpha,d}\,\inf_{\gamma} \int_{0}^1 \frac{\|\gamma'(s)\|_{\mathcal{L}^2}}{f_X\{\gamma(s)\}^{\frac{\alpha-1}{d}}} \,ds \,,
\end{align}
where $\mu_{\alpha,d}$ is the percolation constant that depends only on $\alpha$ and $d$, and $\gamma:[0,1]\to\mathcal{M}$ is a continuously differentiable path satisfying $\gamma(0)=x_1$ and $\gamma(1)=x_2$. Slightly different from standard definition of the Fermat distance, we explicitly include $\mu_{\alpha,d}$ in the definition for notation simplicity in theory. The percolation constant $\mu_{\alpha,d}$ is defined as a limit for $\alpha>1$ \citep{Howard1997}, and we define $\mu_{1,d}=1$ so that $d_{\mathcal{M},1}$ reduces to the geodesic distance $d_{\mathcal{M}}=\inf_{\gamma} \int_{0}^1 \|\gamma'(s)\|_{\mathcal{L}^2} \,ds  $. 

The definition of $d_{\mathcal{M},\alpha}$ at~\eqref{eq_fd} shows that $d_{\mathcal{M},\alpha}$ with $\alpha>1$ favors a path $\gamma$ through a high density region. To illustrate, suppose $d_\mathcal{M}(x_1,x_2) = d_\mathcal{M}(x_1,x_3)$. If $x_1$ and $x_2$ are connected by a path through a cluster (e.g., a dense region of similar spectra), while the paths between $x_1$ and $x_3$ always cross a low-density void, the density $f_X$ in the denominator of~\eqref{eq_fd} ensures that $d_{\mathcal{M},\alpha}(x_1,x_2) < d_{\mathcal{M},\alpha}(x_1,x_3)$. Under the cluster assumption, $x_1$ and $x_2$ are more likely to share the same label. By designing a classifier where a smaller $d_{\mathcal{M},\alpha}$ implies a higher probability of possessing the same label,
we effectively exploit the cluster assumption. Notably, $d_{\mathcal{M},\alpha}$ depends entirely on the marginal distribution of $X$, and thus it can be estimated by taking advantage of the unlabeled sample. 
 
\subsection{Weighted $k$-NN and MDS-based Classifiers}\label{sc_class}
Once pairwise Fermat distances are obtained, they naturally form the foundation for a variety of distance-based classification algorithms. Presumably the simplest one is the $k$-nearest neighbors (NN) classifier, which directly captures the local covariate structure by the metric. To fully exploit the information of the Fermat distance, we propose the following weighted $k$-NN classifier. Specifically, let $\{\pi(1),\ldots,\pi(n_\ell)\}$ be a permutation of $\{1,\ldots,n_\ell\}$
such that $d_{\mathcal{M},\alpha}(X_{\pi(1)},X_0)\leq \ldots\leq d_{\mathcal{M},\alpha}(X_{\pi(n_\ell)},X_0)$, where $X_0$ is a target unlabeled variable. The weighted $k$-NN classifier predicts the label of $X_0$ by
\begin{align}\label{eq_knn_true}
    \tilde{h}_{\textrm{w$k$NN}}(X_0) = \argmax_{y\in \mathcal{Y}}\sum_{i=1}^k w_i \mathds{1}\{Y_{\pi(i)}=y\}\,,
\end{align}
where $\mathds{1}$ denotes the indicator function, and 
\begin{align*}
    w_i = \frac{\exp\{-d_{\mathcal{M},\alpha}(X_{\pi(i)},X_0)/\sigma\}}{\sum_{j=1}^k \exp\{-d_{\mathcal{M},\alpha}(X_{\pi(j)},X_0)/\sigma\}}.
\end{align*} 
Here $\sigma>0$ is a tuning parameter. The unweighted $k$-NN classifier is recovered if $\sigma\to \infty$. The closer individuals in terms of the Fermat distance are assigned with the larger weights, while further individuals contribute less to the classification. The selection of the tuning parameters $k$ and $\sigma$ will be discussed in Section~\ref{sc_class_est}.

Beyond $k$-NN classifiers, we can construct a flexible family of functional classifiers by adapting the multidimensional scaling (MDS) assisted strategy \citep{Tan2025}. This approach projects the functional data into a Euclidean space while preserving pairwise Fermat distances, thereby enabling the use of any given multivariate classifier, say $\tilde{h}$.
Specifically, we apply MDS on the $n\times n$ matrix of pairwise Fermat distances to obtain a sample of multivariate representations $\{X_i^{\rm{MDS}}\}_{i=1}^{n}\subset \mathbb{R}^p$, whose pairwise Euclidean distances approximate the Fermat distances. The classifier $\tilde{h}$ is then trained on $\{X_i^{\rm{MDS}},Y_i\}_{i=1}^{n_\ell}$ and used to predict labels of unlabeled individuals $X_j\in \{X_i\}_{i=n_\ell+1}^{n}$ by $\tilde{h}(X_j^{\rm{MDS}})$. 

In contrast to \citet{Tan2025}, the target dimension $p$ of $\mathbb{R}^p$ here is not necessarily equal to the intrinsic dimension $d$ of the manifold $\mathcal{M}$. In fact, it can be seen as a tuning parameter controlling the level of preserving the pairwise Fermat distances. A perfect (discrete) isometric embedding of a Euclidean distance matrix is ensured by setting $p=n-1$ \citep{Borg2007}. Often, a value of $p<n-1$ is able to sufficiently maintain pairwise Fermat distances in practice. On the other hand, setting $p=d$, the intrinsic dimension or its estimate, is reasonable if the manifold is isometric to a subset of $\mathbb{R}^d$, while choosing $p<d$ may introduce additional geometric distortion. We will evaluate different choices of $p$ on simulated data in Section~\ref{sc_sim_2}. 

The MDS-based functional classifiers form a broad family, because they can use essentially any multivariate classifier. We nevertheless focus on classifiers that exploit distance-related information, such as $k$-NN classifiers and support vector machines (SVM); otherwise, the benefit of the Fermat distance may be masked.
\section{Estimation}\label{sc_est}

We introduce in detail the implementation of the classifiers proposed in Section~\ref{sc_class} on discretely observed functional data. The procedure involves three primary steps: recovering individual trajectories (Section~\ref{sc_fd_est}), estimating the Fermat distance (Section~\ref{sc_fermat_est}), and performing the classification (Section~\ref{sc_class_est}).

\subsection{Functional Trajectory Reconstruction}\label{sc_fd_est}
When the functional observations are discrete and noisy, a presmoothing step is crucial to subsequent analysis. Not only does presmoothing provide functional evaluation on any $t\in[0,1]$, but more importantly, its denoising effect is essential for revealing the low-dimensional intrinsic structure of $X$. Indeed, the observational noise in the ambient space blurs the intrinsic geometric structure of $X$. Recall that a dense design, i.e., the $T_{ij}$'s are dense and distributed on the whole $[0,1]$, is assumed in model~\eqref{eq_discrete}. This dense design, which is satisfied in our astronomical application, is used to control the trajectory-reconstruction error term in our theoretical bounds; see Section \ref{sc_theo}. Based on the dense design, an individually nonparametric smoothing suffices to recover functional trajectories.

Specifically, we adopt the ridged local linear estimator \citep{Lin2021} on individual observations $\{(T_{ij},\tilde{X}_{ij})\}_{j=1}^{J_i}$ under model~\eqref{eq_discrete} to obtain a smooth estimator $\hat{X}_i$ of $X_i$. For $t\in[0,1]$, the standard local linear estimator of $X_i(t)$ is given by $(\mathcal{T}_0\mathcal{S}_2-\mathcal{T}_1\mathcal{S}_1)/(\mathcal{S}_0\mathcal{S}_2-\mathcal{S}_1^2)$, where
\begin{align*}
\mathcal{S}_\ell = \dfrac{1}{J_i} \sum_{j=1}^{J_i} K_{b_i}( T_{ij}-t ) \Big( \dfrac{T_{ij}-t}{b_i}\Big)^\ell\,,~\mathcal{T}_\ell=\dfrac{1}{J_i} \sum_{j=1}^{J_i} K_{b_i}( T_{ij}-t )\Big( \dfrac{T_{ij}-t}{b_i}\Big)^\ell \tilde{X}_{ij}\,,
\end{align*}
for $\ell=0,1$ and $2$. Here, $K_b(\cdot)=K(\cdot/b)/b$ with $K$ the kernel, often a symmetric density, and $b>0$ the bandwidth. A ridge parameter $\Delta$ is introduced to the denominator $(\mathcal{S}_0\mathcal{S}_2-\mathcal{S}_1^2)$ to stabilize the resulting estimator. That is, the ridged local linear estimator of $X_i(t)$ is defined as 
\begin{align}\label{eq_X_est}
\hat{X}_i(t) = \dfrac{\mathcal{T}_0\mathcal{S}_2-\mathcal{T}_1\mathcal{S}_1}{\mathcal{S}_0\mathcal{S}_2-\mathcal{S}_1^2+\Delta \sign(\mathcal{S}_0\mathcal{S}_2-\mathcal{S}_1^2)\mathds{1}\{|\mathcal{S}_0\mathcal{S}_2-\mathcal{S}_1^2|<\Delta\}}\,, ~~\textrm{for } t \in [0,1]\,.
\end{align}
Following \citet{Lin2021}, we set $\Delta=J_i^{-2}$. It is important to note that although $X_i$ lies on the unknown manifold $\mathcal{M}$, its estimator $\hat{X}_i$ does not exactly lie on $\mathcal{M}$ for any finite sample. This poses additional challenges for theoretical investigation and leads to profound effect on classification accuracy under the semi-supervised context.

In the applications of semi-supervised functional classification, the pooled sample size $n$ and the individual sampling rate $J_i$ are often astronomical. Therefore, computation efficiency is required in our estimation procedure. To achieve fast computation in recovering functional trajectories, we select the bandwidth $b_i$ via a plug-in procedure without any cross-validation step; see Section~D of the Supplementary Material for details. 

\subsection{Fermat Distance Estimation}\label{sc_fermat_est}
Estimation of Fermat distance does not require estimation of the density $f_X$. Let $G_{\rm{com}}$ denote the complete graph with all the individuals in the pooled sample $\{X_i\}_{i=1}^n$ as nodes, i.e., any pair $(X_i,X_j)$ with $i\neq j$ is connected in $G_{\rm{com}}$. If $X_i$ were random vectors valued in $\mathbb{R}^D$, the standard approach \citep{Groisman2022,Fernandez2023,Trillos2024} to estimate the Fermat distance is, for $i,j=1,\ldots,n$,
\begin{align*}
\tilde{d}_{\textrm{RV},\alpha} (X_i,X_j) = \min_{\gamma_m\in G_{\textrm{com}}}\sum_{\ell=1}^m \|X_{i_\ell}-X_{i_{\ell+1}}\|_{\mathbb{R}^D}^\alpha\,,
\end{align*} 
where $\gamma_m=(i=i_1,\ldots,i_{m+1}=j)$ is a discrete path in $G_{\textrm{com}}$, i.e., $i_\ell$ and $i_{\ell+1}$ are connected in $G_{\textrm{com}}$, for $\ell = 1,\ldots,m$. A natural idea of extending the approach above to our case is considering 
\begin{align}\label{eq_fd_com}
 \tilde{d}_{\textrm{com},\alpha} (X_i,X_j) = \min_{\gamma_m\in G_{\textrm{com}}}\sum_{\ell=1}^m \|\hat{X}_{i_\ell}-\hat{X}_{i_{\ell+1}}\|_{\mathcal{L}^2}^\alpha\,, 
\end{align}
for $i,j=1,\ldots,n$. For $x_1,x_2\in\mathcal{M}$ but not belonging to $\{X_i\}_{i=1}^n$, $ \tilde{d}_{\textrm{com},\alpha} (x_1,x_2)$ is well defined by fixing the endpoints of $\gamma_m$ as $x_1$ and $x_2$ and considering the complete graph with $\{X_i\}_{i=1}^n\cup\{x_1,x_2\}$ as nodes.

However, finding shortest paths over the complete graph $G_{\rm{com}}$ is computationally heavy for large $n$. A convenient way is restricting to a relatively sparse graph such as a $k_g$-NN graph. Yet, a $k_g$-NN adjacency graph is not necessarily connected, leading to undefined distances between disconnected components. To fix this issue, we propose a simple but computationally efficient remedy by taking the union of a $k_g$-NN graph and the minimum spanning tree (MST) graph as our adjacency graph. That is, we define that two points $X_i$ and $X_j$ are connected in the sparse graph $G_{\rm{sparse}}$ if and only if $\hat{X}_i$ is in the $k_g$-NN of $\hat{X}_j$, or vice versa, or $\hat{X}_i$ and $\hat{X}_j$ are connected in the MST graph. Here, both $k_g$-NN and MST are defined with respect to the $\mathcal{L}^2$ distance between the $\hat{X}_i$'s. This graph construction ensures that every pair of points is path-connected while maintaining graph sparsity for fast path-searching. It follows that we consider
\begin{align*}
 \tilde{d}_{\textrm{sparse},\alpha} (X_i,X_j) = \min_{\gamma_m\in G_{\textrm{sparse}}}\sum_{\ell=1}^m \|\hat{X}_{i_\ell}-\hat{X}_{i_{\ell+1}}\|_{\mathcal{L}^2}^\alpha\,, 
\end{align*}
for $i,j=1,\ldots,n$. The optimization problem above can be efficiently solved by the Dijkstra algorithm.

Finally, based on Theorem~\ref{theo_FD} in Section~\ref{sc_theo}, we apply a sample-size scaling factor to define our estimator of the Fermat distance as 
\begin{align}\label{eq_fd_est}
\hat{d}_{\mathcal{M},\alpha} (X_i,X_j) = n^{\frac{\alpha-1}{d}}\tilde{d}_{\textrm{sparse},\alpha} (X_i,X_j)\,.
\end{align} 
Here, $\alpha>1$ is regarded as a hyperparameter. We suggest pre-specifying a value for $\alpha$ instead of optimizing via a data-driven approach, because the labeled data are usually limited and classification performance is insensitive to a small range larger than 1 as shown in our experiments. The equation~\eqref{eq_fd_est} is also consistent for $\alpha=1$ (but \eqref{eq_fd_com} is not) in the sense that it yields an estimate of the geodesic distance $d_{\mathcal{M}}(X_i,X_j)$ for $\alpha=1$.

If one intends to estimate the Fermat distance between a newly observed $X_{n+1}$ (or its estimate) and any point in the pooled sample $\{X_i\}_{i=1}^{n}$, one needs to construct an adjacency graph $G_{n+1}$ for $n+1$ individuals. To achieve fast computation, we suggest defining $G_{n+1}$ by fixing $G_{\rm{sparse}}$ and connecting $X_{n+1}$ to its nearest $k_g$-NN in $\{X_i\}_{i=1}^{n}$.

\subsection{Semi-supervised Classification}\label{sc_class_est}
Now that we obtain pairwise estimated Fermat distance $\hat{d}_{\mathcal{M},\alpha}(X_i,X_j)$, for $i,j=1,\ldots,n$, we can apply the methodology in Section~\ref{sc_class} to construct classifiers. Our classifiers are semi-supervised in the sense that the $\hat{d}_{\mathcal{M},\alpha}(X_i,X_j)$'s are estimated based on the pooled sample $\{\hat{X}_i\}_{i=1}^n$.

For a target unlabeled observation~$X_0$, let $\{(1),\ldots,(n_\ell)\}$ be a permutation of $\{1,\ldots,n_\ell\}$ such that $\hat{d}_{\mathcal{M},\alpha}(X_{(1)},X_0)\leq \ldots\leq \hat{d}_{\mathcal{M},\alpha}(X_{(n_\ell)},X_0)$. Following~\eqref{eq_knn_true}, the weighted $k$-NN classifier predicts the label of $X_0$ by
\begin{align}\label{eq_knn_est}
\hat{h}_{\textrm{w$k$NN}}(X_0) = \argmax_{y\in \mathcal{Y}}\sum_{i=1}^k \hat{w}_i \mathds{1}\{Y_{(i)}=y\}\,,
\end{align}
where  
\begin{align*}
\hat{w}_i = \frac{\exp\{-\hat{d}_{\mathcal{M},\alpha}(X_{(i)},X_0)/\sigma\}}{\sum_{j=1}^k \exp\{-\hat{d}_{\mathcal{M},\alpha}(X_{(j)},X_0)/\sigma\}}.
\end{align*}
We suggest choosing $\sigma$ via the cross-validation based on the classification accuracy on the labeled sample $\{\hat{X}_i,Y_i\}_{i=1}^{n_\ell}$. The labeled sample size $n_\ell$ is often not large, so this cross validation procedure is not time-consuming. Our theoretical results require $k\asymp [ n_\ell/\log(n_\ell) ]$, where $[\cdot]$ denotes the rounding to the closest integer. In practice, this suggests a simple criterion of setting $k=[n_\ell/n_0]$ with a user-specified $n_0$, say $5$ or $10$.

The MDS-based strategy stated in Section~\ref{sc_class} can be seamlessly applied on the pairwise estimated Fermat distances $\hat{d}_{\mathcal{M},\alpha}(X_i,X_j)$. In our experiments, we will choose SVM as the multivariate classifier $\tilde{h}$ with both linear and Gaussian kernels. Also, we will consider two options of the target dimension $p$ of MDS: (1) $p$ as large as needed for preserving distances; (2) $p=\hat{d}$, where $\hat{d}$ is an estimator of the intrinsic dimension.
 
\section{Theoretical Results}\label{sc_theo}
In theory, it is conventional to consider binary classification for brevity, i.e., $Y\in\{0,1\}$.
For a classifier $h:\mathcal{L}^2([0,1])\to \{0,1\}$, let $\mathcal{R}(h)=P\{Y\neq h(X)\}$ denote the classification risk, and let $\eta (x) = P(Y=1|X=x)$ denote the conditional probability of positive outcome. It is well known that the Bayes classifier $h^*(x) = \mathds{1} \{\eta(x)>1/2\}$ achieves the optimal risk, i.e., $\mathcal{R}(h^*)\leq \mathcal{R}(h)$, for any classifier $h$. A classifier $h$ whose risk converges to $\mathcal{R}(h^*)$ is referred to as an asymptotically optimal classifier. In supervised classification, it is of interest to develop asymptotically optimal classifiers with fast convergence rates. However, it has been shown in \citet{Rigollet2007} that, under the cluster assumption only, the convergence rate of the risk outside the clusters cannot be improved using an additional unlabeled sample, essentially because the cluster assumption only concerns data within clusters. Therefore, instead of the global risk $\mathcal{R}(h)$, we will focus on the risk within clusters. To this end, we first properly define the cluster. Adapting the classical definition of cluster as a connected region where the density is high, we define the cluster with respect to the Fermat distance as follows.
\begin{defi}[Cluster]
For fixed $\alpha \geq 1$ and $\kappa>0$, a closed subset $T\subset \mathcal{M}$ is a cluster with respect to $\alpha$ and $\kappa$, if $\forall x_1,x_2\in T$, there exists an optimal path $\gamma_{12}$ found by the power-$\alpha$ Fermat distance $d_{\mathcal{M},\alpha}(x_1,x_2)$ satisfying $\gamma_{12}(t)\in T$ and $f_{X} \{\gamma_{12}(t)\}\geq \kappa$,  $\forall t\in[0,1]$.
\end{defi}

We assume that there is a set of disjoint clusters $\{U_j\}_{j=1}^{n_c}\subset \mathcal{M}$, which are used to exposit the theoretical properties of classifiers. For fixed $\alpha\geq 1$ and $\kappa>0$, the sets $\{U_j\}_{j=1}^{n_c}$ are fixed but unknown subsets of $\mathcal{M}$. The risk within clusters is then defined as $\mathcal{R}_{\mathcal{T}}(h) = P(Y\neq h(X),X\in \cup_{j=1}^{n_c} U_j)$, for any classifier $h$. Since we will consider both the supervised and semi-supervised cases, we use $E_{n_\ell}(\cdot)$ to denote the expectation with respect to the labeled sample. Similarly, $E_{n}(\cdot)$ denotes the expectation with respect to the pooled sample, $P_{n_\ell}(\cdot)$ and $P_n(\cdot)$ denote the corresponding probabilities, while $P(\cdot)$ and $E(\cdot)$ denote the probability and expectation that do not depend on sample size. In this section as well as the Supplementary Material, we use $C$ to denote a generic positive constant that may depend on fixed parameters and vary under different contexts.

Because the MDS-based classifiers depend on a specific multivariate classifier, we focus on the weighted $k$-NN classifier. 
The following assumptions are needed for the classification property of the weighted $k$-NN classifier with respect to the true Fermat distance. 

\paragraph{Assumption A}
\begin{itemize}
\item[\namedlabel{CA1}{(A1)}] There exists an $\epsilon_0>0$ such that $\inf_{x\in\cup_{j=1}^{n_c} U_j}|\eta(x)-1/2|>\epsilon_0$.

\item[\namedlabel{CA2}{(A2)}] The conditional probability $\eta(\cdot)$ is Lipschitz continuous with respect to the power-$\alpha$ Fermat distance, i.e., $\exists L>0$ such that $\forall x_1,x_2\in \mathcal{M}$, $|\eta(x_1)-\eta(x_2)|\leq Ld_{\mathcal{M},\alpha}(x_1,x_2)$.

\item[\namedlabel{CA3}{(A3)}] The density $f_X$ is continuously differentiable with $\inf_{x\in\mathcal{M}}f_X(x)>0$. 
  
\end{itemize}
The assumption~\ref{CA1} is referred to as the cluster assumption, which implies that, for all $U_j$,
\begin{align*}
P(Y_1=Y_2\mid X_1,X_2\in U_j)\geq P(Y_1\neq Y_2\mid X_1,X_2\in U_j)\,.
\end{align*}
This inequality directly reflects the role of clusters in task of classification. The assumption~\ref{CA1} is slightly stronger than the cluster assumption defined in \citet{Rigollet2007}, which requires that $\mathds{1}\{\eta(x)>1/2\}$ is a constant on any given cluster. Here, we require a strict gap for more tractable theoretical development. In classical theory of supervised classification \citep{Massart2006}, the following assumption has often been invoked: For some $\epsilon_0>0$,
$ |\eta(x)-1/2|>\epsilon_0$. This assumption is clearly stronger than the cluster assumption \ref{CA1}, because we only require such a gap to exist on clusters.  Assumption~\ref{CA2} is the Lipschitz condition adapted to the Fermat distance, which is standard in the literature on the classification theory \citep{Audibert2007,Gadat2016}. Assumption~\ref{CA3} is a mild condition to ensure the Fermat distance is well defined for any pair of points on $\mathcal{M}$ \citep{Hwang2016}; this assumption is also used to provide a lower bound for $P(d_{\mathcal{M},\alpha}(X,x)\leq t)$, for a small $t>0$. 

\begin{theo}\label{theo_upper}
Under Assumptions~\ref{CA1} to \ref{CA3}, set $k\asymp [ n_\ell/\log(n_\ell) ]$ and $\sigma\in[\sigma_0,\sigma_1] $ for some fixed and positive $\sigma_0$ and $\sigma_1$, then there exists a constant $C>0$ and $N_{\ell0} \in \mathbb{N}^+$ such that for all $n_\ell>N_{\ell0}$,
\begin{align*}
    E_{n_\ell}\{\mathcal{R}_{\mathcal{T}}(\tilde{h}_{\textrm{w$k$NN}})\}-\mathcal{R}_\mathcal{T} (h^*)\leq \exp\Big\{-C\frac{n_\ell}{\log(n_\ell)}\Big\}\,.
\end{align*}
\end{theo}
The proof of Theorem~\ref{theo_upper} is given in Section~A of the Supplementary Material. Theorem~\ref{theo_upper} shows that the expected excess risk within clusters of the weighted $k$-NN classifier using the true Fermat distance is exponentially decaying with respect to the labeled sample size. Such a fast rate is mainly due to the cluster assumption~\ref{CA1}, based on which the bias of $\tilde{h}_{\textrm{w$k$NN}}$ is sufficiently small. We can then choose a larger $k$ to reduce the variance of $\tilde{h}_{\textrm{w$k$NN}}$, which leads to the final fast rate. Specifically, $k\asymp [ n_\ell/\log(n_\ell) ]$ is of larger order than the optimal choice $k\asymp [ n_\ell^{2/(2+d)} ]$ with respect to the classical risk \citep{Gadat2016}, where $[\cdot]$ denotes the rounding to the closest integer.

Next, we consider the weighted $k$-NN classifier using the estimated Fermat distance. We first provide a consistency result for the estimated Fermat distance. A notable distinction from the existing literature \citep{Hwang2016,Groisman2022,Fernandez2023} is that here we only observe a discrete and noisy version of the functional data. It is inevitable to take into account the estimation error of recovering the functional trajectories. This error plays a vital role in explaining semi-supervised classification performance.

Under the discrete observation model~\eqref{eq_discrete}, we assume the same sampling rate for all individuals for the sake of notation simplicity, i.e., $J_i \asymp J$, and thus we can use a single bandwidth $b_i\equiv b $ in the ridged local linear estimator.
Recall from~\eqref{eq_fd_est} that $\hat{d}_{\mathcal{M},\alpha}$ is defined on the sparse graph $G_{\rm{sparse}}$ for computational efficiency. We restrict the theoretical analysis to the complete graph. That is, we consider $\hat{d}_{\mathcal{M},\alpha} (X_i,X_j) = n^{\frac{\alpha-1}{d}}\tilde{d}_{\textrm{com},\alpha} (X_i,X_j)$ with $\tilde{d}_{\textrm{com},\alpha}$ given in \eqref{eq_fd_com}. It is still possible to consider a $k_g$-NN graph by choosing a properly growing $k_g$ \citep{Groisman2022,Little2022}, but such theoretical work deviates from our main concern. In experiments, we will use a relatively large $k_g$ to be more aligned with the theory as well as compare the classification accuracy using either $G_{\rm{sparse}}$ or $G_{\rm{com}}$. The following assumptions concerning model~\eqref{eq_discrete} are needed for consistency of the estimated Fermat distance.
\paragraph{Assumption B}

\begin{itemize} 

\item[\namedlabel{CB1}{(B1)}] The design points $T_{ij}$ are fixed and uniform, i.e., $T_{ij}=j/J$, for $i=1,\ldots,n$, $j=1,\ldots,J$. 

\item[\namedlabel{CB2}{(B2)}] The functional variable $X_i$ is twice continuously differentiable with $P(\sup_{t\in[0,1]}\|X_i^{(2)}(t)\|\leq C)=1$, for some constant $C>0$. 

\item[\namedlabel{CB3}{(B3)}] The error variable $\epsilon_{ij}$ is sub-Gaussian, i.e., $\forall t>0$, $P(|\epsilon_{ij}|\geq t)\leq 2\exp(-t^2/C^2) $, for some constant $C>0$. 

\item[\namedlabel{CB4}{(B4)}] The kernel $K$ is a continuously differentiable and symmetric density supported on $[-1,1]$ satisfying $\int_{-1}^1 K^{(\ell)}(u)u^2\,du <\infty$, and $\int_{-1}^1 K^2(u)\,du<\infty$, for $\ell=0$ and $1$.  
 
\end{itemize}
Assumptions~\ref{CB1} to \ref{CB4} are mild regularity assumptions to show that the tail probability of the ridged local linear estimator is exponentially decaying. Their similar variants are commonly assumed in the literature of local polynomial smoothing \citep{Fan1996,Lin2021}. The fixed and uniform design required in  \ref{CB1} is not too restrictive, as classical results are stated conditional on design points if they are random. Here, we require the fixed and uniform design (satisfied in our real data application) to simplify theoretical exposition. 

\begin{theo}\label{theo_FD}
Under Assumptions~\ref{CA3}, \ref{CB1} to \ref{CB4}, fix $\alpha>1$, if $b^2n^{1/d}\to 0$ and $J bn^{-2/d}/\log (nJ)\to \infty$, then $\forall \epsilon>0$, $\exists C_\epsilon>0$, $\exists N_\epsilon\in\mathbb{N}^+$ such that for all $n>N_\epsilon$, 
\begin{align*}
P_n\Big(\sup_{x_1,x_2\in\mathcal{M}}\Big|\hat{d}_{\mathcal{M},\alpha} (x_1,x_2)-d_{\mathcal{M},\alpha}(x_1,x_2) \Big|>\epsilon\Big) \leq \exp(-C_\epsilon n^{\frac{1-\lambda  }{d+2\alpha} }) +  \exp(-CJbn^{-2/d}\epsilon^2 )\,, 
\end{align*}
for some $\lambda\in((\alpha-1)/\alpha,1)$ and some constant $C>0$. 
\end{theo}

The proof of Theorem~\ref{theo_FD} is given in Section~B of the Supplementary Material. Theorem~\ref{theo_FD} shows that the uniform tail probability of the estimated Fermat distance is exponentially decaying. More specifically, the rate contains two terms: The first term $\exp(-C_\epsilon n^{\frac{1-\lambda }{ d+2\alpha} })$ is the same as the one in the proposition 5 of \citet{Fernandez2023}, which states a similar result for noise-free random vectors. The second term $\exp(-C_1Jbn^{-2/d}\epsilon^2 )$ originates from estimation of the functional data from discrete and noisy observations. The estimation error of the functional data is aggregated in the graph $G_{\rm{com}}$ so that the bandwidth order and the individual sampling rate required for consistency are $b^2n^{1/d}\to 0$ and $J bn^{-2/d}/\log (nJ)\to \infty$. Intuitively, the graph-aggregated bias is of order $O(b^2n^{1/d})$ and the corresponding variance is of order  $O\{n^{2/d}\log (nJ)/(J b)\}$. For comparison, it is well known that the bias and variance of the local linear estimator are of order $O(b^2)$ and $O\{\log (J)/(J b)\}$, respectively \citep{Fan1996}. It is worth noting that, in our case, the optimal bandwidth order based on the bias-variance trade-off remains $O(J^{-1/5})$ up to a logarithmic factor, which is satisfied by our plug-in criterion detailed in Section~D of the Supplementary Material. 

The discussion above shows the individual sampling rate such that the local linear smoothing is consistent is not sufficient for the estimated Fermat distance to be consistent. Specifically, in a sufficiently dense design where $Jbn^{-\frac{(3-\lambda) d +4\alpha}{ d(d+2\alpha)}} \gtrsim 1$, we have $\exp(-CJbn^{-2/d}\epsilon^2 ) = O\{\exp(-C_\epsilon n^{\frac{1-\lambda  }{d+2\alpha} })\}$, and thus the rate reduces to the classical rate as in the proposition 5 of \citet{Fernandez2023} as if the true $X_i$'s are available. On the contrary, in the case where $Jbn^{-\frac{(3-\lambda) d +4\alpha}{ d(d+2\alpha)}} \to 0$, the error from recovering functional trajectories dominates.

Now combining Theorems~\ref{theo_upper} and \ref{theo_FD}, we are able to provide the convergence rate of the expected risk within clusters of the weighted $k$-NN classifier using the estimated Fermat distance.

\begin{theo}\label{theo_upper_est}
Under Assumptions~\ref{CA1} to \ref{CA3}, \ref{CB1} to \ref{CB4}, fix $\alpha>1$, set $k\asymp [ n_\ell/\log(n_\ell) ]$ and $\sigma\in[\sigma_0,\sigma_1] $ for some fixed and positive $\sigma_0$ and $\sigma_1$, if $b^2n^{1/d}\to 0$ and $J bn^{-2/d}/\log (nJ)\to \infty$, then there exist constants $ C>0$, 
$N_{\ell0}\in\mathbb{N}^+$ and $N\in\mathbb{N}^+$ such that
$\forall n >N$ and $\forall n_\ell>N_{\ell0}$, 
\begin{align*} 
&E_{n}\{\mathcal{R}_\mathcal{T}(\hat{h}_{\textrm{w$k$NN}})\}-\mathcal{R}_\mathcal{T}(h^*) \\
&\leq \exp\Big\{- C\frac{n_\ell}{\log(n_\ell)} \Big\}+  \exp(-C n^{\frac{1-\lambda  }{d+2\alpha} }) + \exp(-C Jbn^{-2/d} )\,.
\end{align*}   
\end{theo}

The proof of Theorem~\ref{theo_upper_est} is given in Section~C of the Supplementary Material. We see that the convergence rate of $E_{n}\{\mathcal{R}_\mathcal{T}(\hat{h}_{\textrm{w$k$NN}})\}-\mathcal{R}_\mathcal{T}(h^*)$ contains three terms: the first one is the same as that from Theorem~\ref{theo_upper} depending on the labeled sample size; the second and third terms originate from estimating the Fermat distance. Intuitively, one may expect that a larger unlabeled sample size would always reduce the error from estimating the Fermat distance. However, the second and third terms tell that increasing the pooled sample improves the noise-free geometric approximation but simultaneously requires more accurate trajectory reconstruction.
When $Jbn^{-\frac{(3-\lambda) d +4\alpha}{ d(d+2\alpha)}} \gtrsim 1$ so that $\exp(-C_\delta'Jbn^{-2/d} )=O\{\exp(-C_\delta n^{\frac{1-\lambda  }{d+2\alpha} }) \}$, together with the common situation where $n_u\gg n_\ell$, the convergence rate of $E_{n}\{\mathcal{R}_\mathcal{T}(\hat{h}_{\textrm{w$k$NN}})\}-\mathcal{R}_\mathcal{T}(h^*)$ is as fast as if the true Fermat distance is used.  

On the other hand, Theorem~\ref{theo_upper_est} suggests that in the case of sparse or moderate design, the unlabeled sample may not be beneficial in terms of improving the convergence rate of the expected excess risk within clusters. More specifically, we summarize the findings as follows:
\begin{itemize}
    \item The error of recovering functional trajectories is negligible in the vanishing upper bound if $Jbn^{-\frac{(3-\lambda) d +4\alpha}{ d(d+2\alpha)}} \gtrsim 1$;
    \item The error of recovering functional trajectories dominates in the vanishing upper bound if $Jbn^{-\frac{(3-\lambda) d +4\alpha}{ d(d+2\alpha)}} \to 0$ and $Jb n^{-2/d} \to \infty$;
    \item The upper bound is not vanishing and thus the unlabeled sample is not guaranteed to be beneficial if $Jb n^{-2/d}$ fails to diverge.
\end{itemize}
Although the theoretical results based on upper bounds are not sufficient to prove the unlabeled sample is useless under the sparse or moderate design, we empirically show by simulated examples that the unlabeled sample improves classification accuracy only when the sampling rate $J$ is growing for a few simulated examples in Section~\ref{sc_sim_2}. 

\section{Simulation Studies}\label{sc_sim}
\subsection{Benchmark Comparison}\label{sc_sim_1}
We evaluate on simulated data classification performance of our semi-supervised classifiers compared with existing supervised classifiers. Specifically, let FD-w$k$NN be the weighted $k$-NN classifier defined at \eqref{eq_knn_est}. In computing an MDS-based classifier, we need to select the target dimension $p$, which controls how well the estimated pairwise Fermat distances are preserved. We consider two choices: (1) $p$ as large as needed to better preserve
the pairwise distances; (2) $p=\hat{d}$, the estimated intrinsic dimension. We use $^{\textrm{high}-p}$ and $^{\textrm{low}-p}$ to denote the choices of (1) and (2), respectively. We choose SVM as the multivariate classifier including both linear and Gaussian kernels. Together with the two choices of the target dimension, there are four MDS-based classifiers, namely FD-LSVM$^{\textrm{high}-p}$, FD-LSVM$^{\textrm{low}-p}$, FD-GSVM$^{\textrm{high}-p}$, and FD-GSVM$^{\textrm{low}-p}$. In this subsection, we focus on FD-GSVM$^{\textrm{low}-p}$, and we will consider other MDS-based classifiers in Section~\ref{sc_sim_2}. Regarding supervised classifiers for comparison, we consider the vanilla $k$-NN classifier based on the $\mathcal{L}^2$ distance (naive-$k$NN), the nonparametric Bayesian classifier with the kernel density estimate \citep[NB,][]{Dai2017}, and the functional supervised manifold learning approach with the $k$-NN classifier \citep[FSML-$k$NN,][]{Tan2025}. NB and FSML-$k$NN are shown to be asymptotically optimal as the labeled sample size tends to infinity. Also, it is interesting to compare different functional variants of the $k$-NN classifier.

Letting $t_1,\ldots,t_J$ be equidistant points in $[0,1]$, we generate data from the following models:
\begin{itemize}
    \item[(i)] $X_i(t_j)=\mu_i(t_j) = -Z_{i1}\exp\{-(t_j-0.2)^2/0.005\}-Z_{i2}\exp\{-(t_j-0.5)^2/0.005\}-Z_{i1}Z_{i2}\exp\{-(t_j-0.8)^2/0.005\}/5$, where $(Z_{i1}|Y_i=y)=\{\theta_i\cos(\theta_i+2y\pi/3)+\delta_i+15\}/10$ and $(Z_{i2}|Y_i=y)=\{\theta_i\sin(\theta_i+2y\pi/3)+\delta_i+15\}/10$. Here, $\theta_i\overset{iid}{\sim}U[1,2\pi]$, and $\delta_i\overset{iid}{\sim}N(0,0.5^2)$.

    \item[(ii)] $X_i(t_j)=\mu_i\{\gamma_i(t_j)\}$, where $\mu_i$ is the same as in model (i) and $\gamma_i(t)=\{\exp(\beta_{i}t)-1\}/\{\exp(\beta_i)-1\}$ with $\beta_i\overset{iid}{\sim}U[-0.5,0.5]$. Set $\gamma_i(t)=t$ for $\beta_i=0$.

    \item[(iii)] $X_i(t_j)=\mu_i\{\gamma_i(t_j)\}$, where $\mu_i$ is the same as in model (i) and $\gamma_i(t)=\{\exp(\beta_{i}t)-1\}/\{\exp(\beta_i)-1\}$ with $\beta_i|Y_i=0\overset{iid}{\sim}U[-0.5,0]$, $\beta_i|Y_i=1\overset{iid}{\sim}U[-0.25,0.25]$, and $\beta_i|Y_i=2\overset{iid}{\sim}U[0,0.5]$. Set $\gamma_i(t)=t$ for $\beta_i=0$.

    \item[(iv)] $X_i(t_j) =  \mathds{1}\{Y_i=1\} \sin(4\pi t_j) + \sum_{\ell=1}^{50} Z_{i\ell} \phi_\ell(t_j)$, where $\phi_1(t_j)=1,\phi_{2\ell+1}(t_j)=\sqrt{2}\sin(2\ell\pi t_j)$,  $\phi_{2\ell}(t_j)=\sqrt{2}\cos(2\ell\pi t_j)$, and $Z_{i\ell}|Y_i=y\sim N(0,\sigma_{\ell y}^2)$ with $\sigma_{\ell0} = \exp(-\ell/6)$ and $\sigma_{\ell1} = \exp(-\ell/4)$.
\end{itemize}

There are three classes (i.e., $Y_i\in\{0,1,2\}$) under models (i) to (iii), and two classes (i.e., $Y_i\in\{0,1\}$) under model (iv). The distribution of $Y_i$ is the discrete uniform distribution for all models. Recall that our real data example consists of astronomical spectra. The function $\mu_i$ under models (i) to (iii) mimics the shape of an astronomical spectrum by a mixture of three Gaussian densities. Moreover, the randomness of $\mu_i$ originates from $Z_i$, whose distribution given $Y_i=0,1$ and $2$ forms three spirals that are commonly used as synthetic data in the literature of semi-supervised learning \citep{Zhu2003,Belkin2006}. Models (ii) and (iii) are time-warping variants of model (i), where the warping function $\gamma_i$ does not (resp., does) contain information of classes under model (ii) (resp., model (iii)). Note that the time warping, which occurs in an astronomical spectrum due to the Doppler effect, is a common source of nonlinearity in functional data \citep{Ramsay2005,Chen2012}. Therefore, models (i) to (iii) are proper data-generating models that encode the manifold and cluster assumptions and yield data similar to our real data example. Finally, model (iv) is a standard Gaussian model slightly modified from \citet{Dai2017}. It is used to test classification performance of the semi-supervised classifiers when the  cluster assumption fails.

\begin{figure}[t]
\centering
\includegraphics[width=0.85\textwidth]{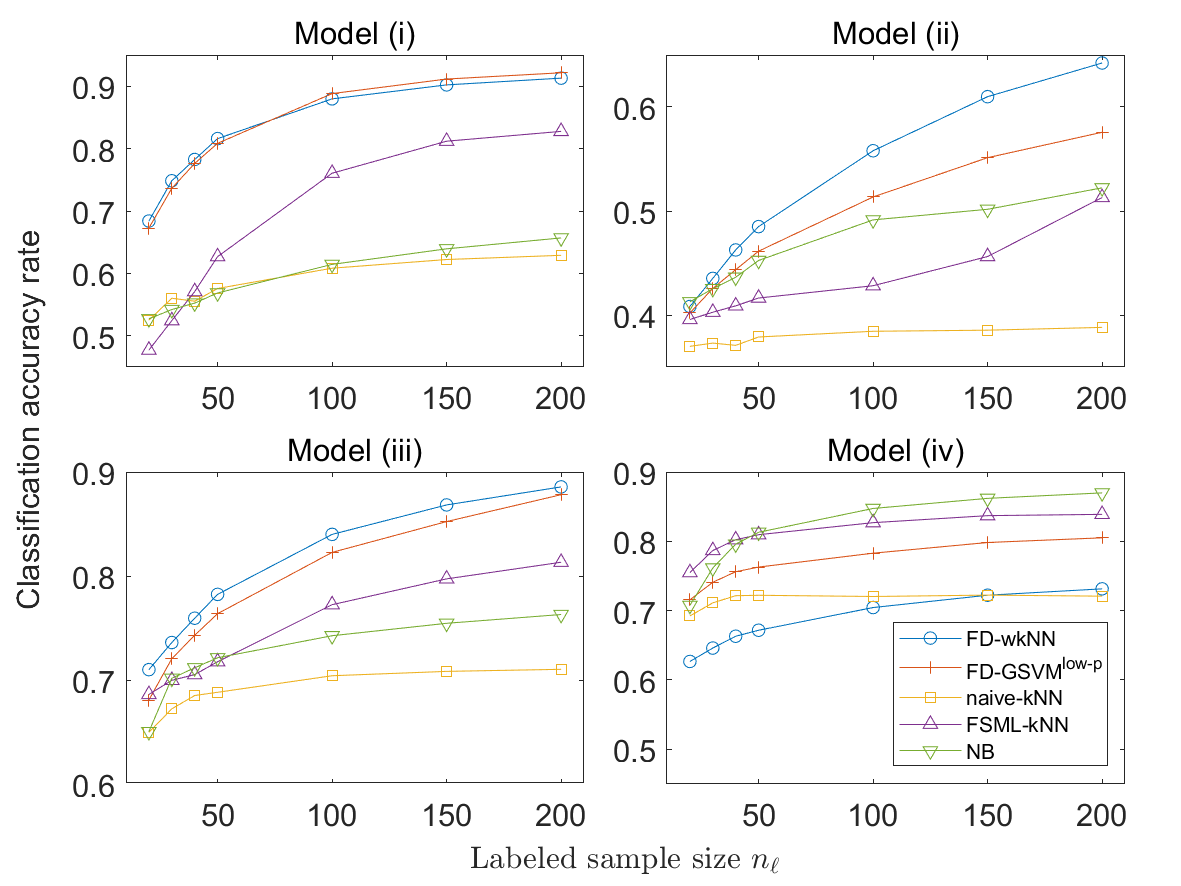} 
\caption{The average rates of classification accuracy of all classifiers under simulation models (i) to (iv) with different labeled sample sizes.}\label{fg_sim_bench}
\end{figure}

We fix the pooled sample size $n=n_\ell+n_u=1000$ and the individual sampling rate $J=100$. All the considered values for $n_\ell$ include $20, 30, 40, 50, 100, 150 $, and $200$. Our observed data consist of $\{(t_j,\tilde{X}_{ij},Y_i)_{j=1}^J\}_{i=1}^{n_\ell}$ and $\{(t_j,\tilde{X}_{ij})_{j=1}^J\}_{i=n_\ell+1}^{n}$, where $\tilde{X}_{ij}=X_i(t_j)+\epsilon_{ij}$ with $\epsilon_{ij}\overset{iid}{\sim}N(0,\sigma_x^2)$. Here, $\sigma_x^2=\hat{V}_x/R$, where $\hat{V}_x$ is the empirical variance of $X_i$ integrated over $[0,1]$, $R=4$ for models (i) to (iii) and $R=20$ for models (iv). The discrete and noisy data $\{(t_j,\tilde{X}_{ij})_{j=1}^J\}_{i=1}^{n}$ are presmoothed using the ridged local linear estimator as in Section~\ref{sc_fd_est} before implementing any classifier. To compute the Fermat distance, we utilize union of $100$-NN graph and the MST graph as the adjacency graph, and set $\alpha=2$. The intrinsic dimension $d$ is estimated using the approach from \citet{Tan2024}. The value of $k$ is set as $[n_\ell/5]$ for all $k$-NN related classifiers.

For each of the models above, we generate 100 datasets and implement all the classifiers to predict the labels of the unlabeled sample. The average rates of classification accuracy under all settings are shown in Figure~\ref{fg_sim_bench}. We see that, under models (i) to (iii) where the manifold and cluster assumptions hold, the semi-supervised classifiers uniformly outperform the others under nearly all settings, except that NB is competitive under model (ii) with $n_\ell\leq 50$. This demonstrates effectiveness of our methodology. More specifically, under models (ii) to (iii), FD-w$k$NN performs slightly better than FD-GSVM$^{\textrm{low}-p}$, while these two perform quite closely under model (i). The performances of FSML-$k$NN and NB are mixed and both are better than that of naive-$k$NN under models (i) to (iii).
Under the Gaussian model (iv), NB and FSML-$k$NN perform similarly and better than the FD-related classifiers. This is reasonable, because the optimal class boundary of model (iv) does not lie on a low-density region, i.e., the cluster assumption fails, which renders the Fermat distance less useful.

\subsection{Sensitivity Analysis of Semi-supervised Classifiers}\label{sc_sim_2}

In this subsection, we evaluate classification performance within the semi-supervised classifiers under different tuning parameters and model settings. We focus on models (i) to (iii) as in Section~\ref{sc_sim_1} where the manifold and cluster assumptions hold.

\begin{figure}[t]
\centering
\includegraphics[width=1\textwidth]{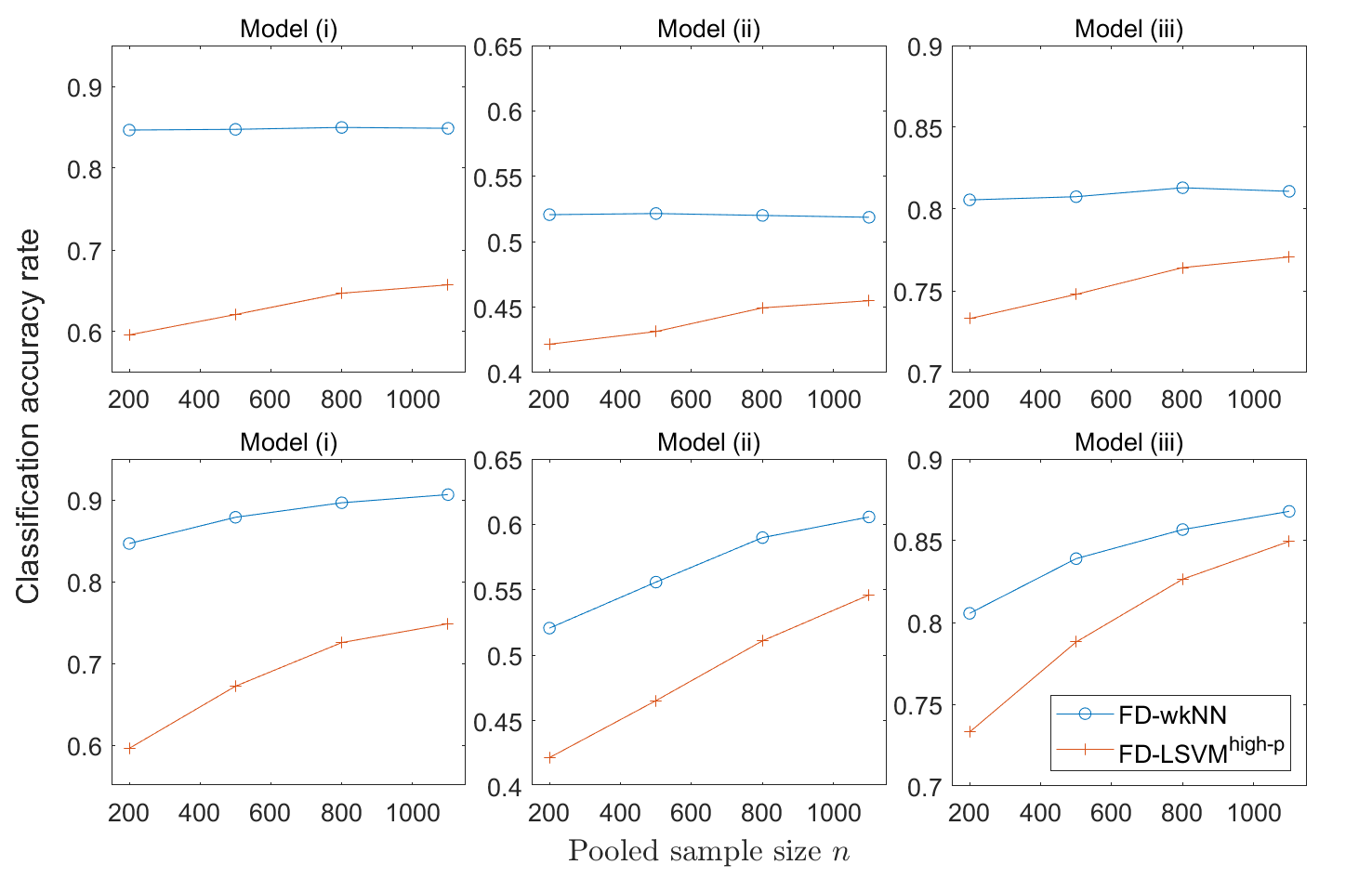} 
\caption{The average rates of classification accuracy of semi-supervised classifiers with $J\equiv 50$ (first row) and $J=(50,100,200,400)$  (second row) and different sample sizes under models (i) to (iii). }\label{fg_sim_diff_sampling}
\end{figure}

We first examine how the effect of increasing the unlabeled sample depends on the individual sampling rate, i.e., empirically supporting Theorem~\ref{theo_upper_est}. Specifically, we fix the labeled sample size $n_\ell=100$ and consider two sets of sampling options: (1) $(J,n)=(50,200),(50,500), (50,800)$ and $(50,1100)$; (2) $(J,n)=(50,200),(100,500), (200,800)$ and $(400,1100)$. While the unlabeled sample size $n_u$ grows under both options, the individual sampling rate $J$ increases only under the option (2). The other model settings are the same as in Section~\ref{sc_sim_1}. We apply FD-w$k$NN and FD-LSVM$^{\textrm{high}-p}$ on 100 datasets generated from each of models (i) to (iii) and show the average rates of classification accuracy in Figure~\ref{fg_sim_diff_sampling}. We see from Figure~\ref{fg_sim_diff_sampling} that the classification accuracy increases significantly with larger unlabeled sample sizes only when the individual sampling rate $J$ also grows. If the individual sampling rate is fixed at a low level, i.e., $J=50$, the classification performances remain similar or slightly better despite a growing unlabeled sample. Such empirical results are consistent with the reconstruction-error mechanism identified in Theorem~\ref{theo_upper_est}.

\begin{figure}[t]
\centering
\includegraphics[width=1\textwidth]{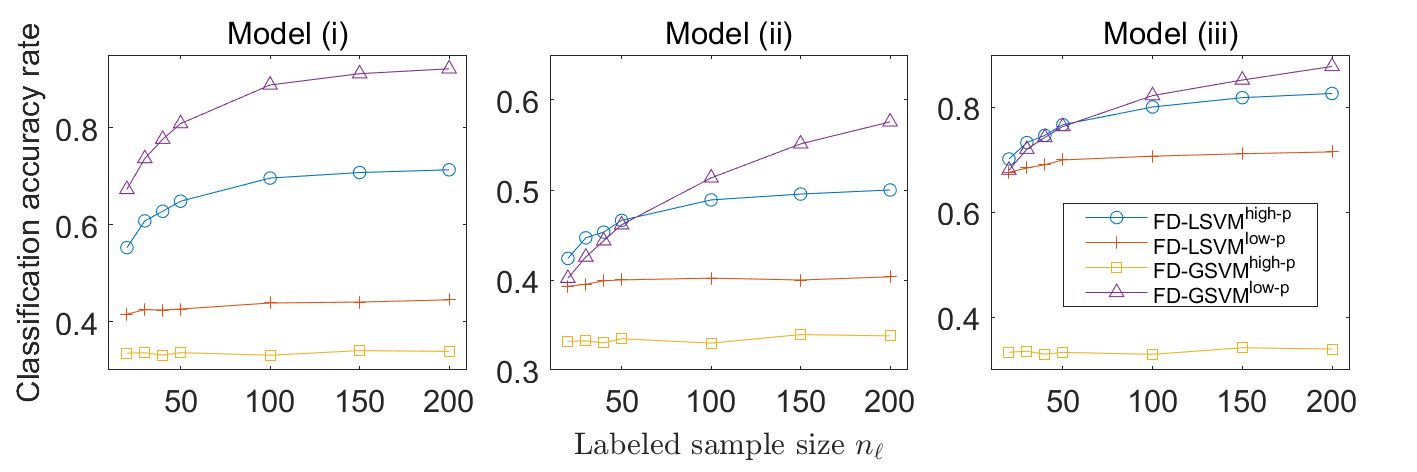} 
\caption{The average rates of classification accuracy of SVM-related classifiers under simulation models (i) to (iii) with different labeled sample sizes.}\label{fg_sim_diff_SVM}
\end{figure}

Next, we test different choices used in MDS-based classifiers. We follow the procedure in Section~\ref{sc_sim_1} to evaluate these classifiers and show the results in Figure~\ref{fg_sim_diff_SVM}. We see from Figure~\ref{fg_sim_diff_SVM} that FD-GSVM$^{\textrm{low}-p}$ performs the best overall, followed by FD-LSVM$^{\textrm{high}-p}$, while the other two perform less well. The result that FD-LSVM$^{\textrm{high}-p}$ outperforms FD-LSVM$^{\textrm{low}-p}$ aligns with the consensus that a linear classifier often benefits when the data lying on a low-dimensional space are lifted to a higher-dimensional space \citep{Hastie2009}. On the other hand, using a Gaussian kernel essentially lifts the data to an infinite-dimensional space, which blurs distance information if the data already lie on a high-dimensional space. This is why FD-GSVM$^{\textrm{high}-p}$ performs quite poorly. In conclusion, for SVM used in MDS-based classifiers, we recommend using either the linear kernel with a high target dimension or the Gaussian kernel with a low target dimension.

As another simulation study, we evaluate the sensitivity to $\alpha$ in computing the Fermat distance. Specifically, we follow the same procedure as in Section~\ref{sc_sim_1} to implement FD-w$k$NN and FD-GSVM$^{\textrm{low}-p}$ with $\alpha=1,2,4,$ and 8. Note that the Fermat distance (both true and estimated) reduces to the geodesic distance for $\alpha=1$. The results are plotted in Figure~S1 of Section E of the supplementary material. We see that overall $\alpha=2$ leads to the best results while $\alpha=8$ yields the worst results, but the differences are tiny. This suggests that the Fermat-distance-induced classifiers are robust to the value of $\alpha$ for a small range of $\alpha\geq 1$.

Moreover, we compare the classification accuracy of three FD-related classifiers using either the complete adjacency graph or the union of the $100$-NN graph and the MST graph. The results are reported in Section~E of the Supplementary Material. We see that these two choices lead to tiny relative differences (up to $0.5\%$) in the classification accuracy, showing that FD-related classifiers are robust to the construction of an adjacency graph under the considered simulation settings.

Finally, we investigate the computational cost of our semi-supervised classifiers. The detailed settings and results are reported in Section~E of the Supplementary Material. In summary, for a pooled sample size around 1000, our classification algorithm only takes two seconds; for $n=8000$, it requires roughly four minutes. This scalability ensures the methodology remains highly practical for modern large-scale functional datasets.

\section{Application to Astronomical Spectral Data}\label{sc_real}

\subsection{Data Description and Preprocessing}\label{sc_real_des}

Gaia is a space astrometry mission of the European Space Agency (ESA) designed to map the Milky Way in six-dimensional (location and velocity) phase space and to provide a wide range of complementary astrophysical data products \citep{GaiaCollaboration2016,GaiaCollaboration2023}. In this application, we use the Gaia DR3 radial velocity spectrometer (RVS) mean spectra, which are provided in the 8460--8700\,$\mathring{A}$ (angstrom) wavelength window. This spectral window contains the Ca II triplet, a prominent set of lines widely used in stellar spectroscopy and kinematic studies.

Our dataset is constructed by joining the Gaia DR3 tables \texttt{vari\_classifier\_result}, \texttt{gaia\_source}, and \texttt{rvs\_mean\_spectrum} through \texttt{source\_id}. 
Only sources with available RVS mean spectra and \texttt{rvs\_spec\_sig\_to\_noise} $\geq 80$ are retained for further analysis. To assign variability classes to Gaia sources, \citet{Rimoldini2023} developed a classifier based on distributed random forest and XGBoost. The training data consist of an extensive cross-match of Gaia sources with catalogs of known stars from existing literature. The estimated classes stored in \texttt{best\_class\_name} correspond to those with the highest scores from \texttt{best\_class\_score}. There, the highest scores are obtained by carefully calibrated posterior probabilities, which form a measure of classification confidence. Since these class assignments are estimated rather than expert-labeled, we later use the scores to distinguish relatively reliable labels from uncertain ones in our semi-supervised learning.

We consider a four-class classification problem using Gaia DR3 RVS mean spectra: a merged DSCT-group class (combining Delta Scuti, Gamma Doradus, and SX Phoenicis candidates), SOLAR\_LIKE, ECL, and YSO. This is a scientifically compelling yet challenging task. For example, identifying YSO from the others is relatively easy, because the Calcium lines often flip into emission, while that of other types of stars show absorption. On the other hand, DSCT and ECL often look similar blurring the boundary of these two classes. Moreover, the number of DSCT and SOLAR\_LIKE is generally much larger than that of ECL and YSO, which properly reflect the phenomenon of imbalanced classes that is commonly seen in astronomic data. Specifically, the total number of spectra is 2981, including 1064 DSCT, 1613 SOLAR\_LIKE, 257 ECL, and 47 YSO.



Missing values in Gaia DR3 RVS spectra can arise because some wavelength pixels are masked at the CCD-sample level, a problem reported to occur more often near the spectral edges \citep{RecioBlanco2023}. After truncating a small proportion of missing values near boundaries of the 8460–8700 $\mathring{A}$ window, each spectrum consists of $J=2221$ pairs of equidistant observations $(T_{ij},\tilde{X}_{ij})$, i.e., a dense and uniform design. The flux values $\tilde{X}_{ij}$ have been normalized to around one. The observations of flux error are used to estimate the variance of $\epsilon_{ij}$ to select the bandwidth used in the ridged local linear smoothing. Specifically, we follow the procedure in Section~\ref{sc_fd_est} with the bandwidth selection detailed in Section D of the Supplementary Material to obtain the smoothed spectra $\hat{X}_i$.

\subsection{Semi-supervised Classification}
 
\begin{figure}[t]
\centering
\includegraphics[width=0.48\textwidth]{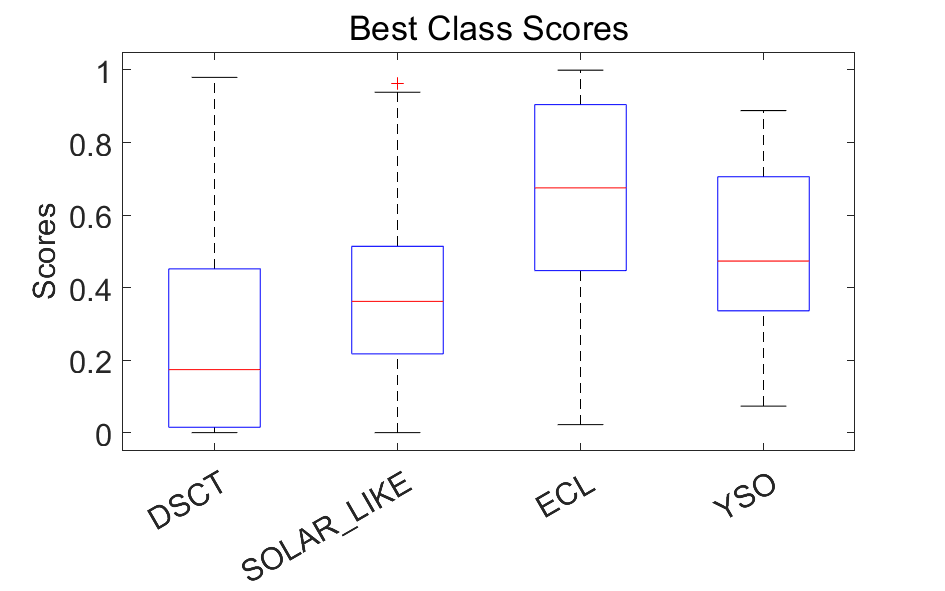}
\includegraphics[width=0.48\textwidth]{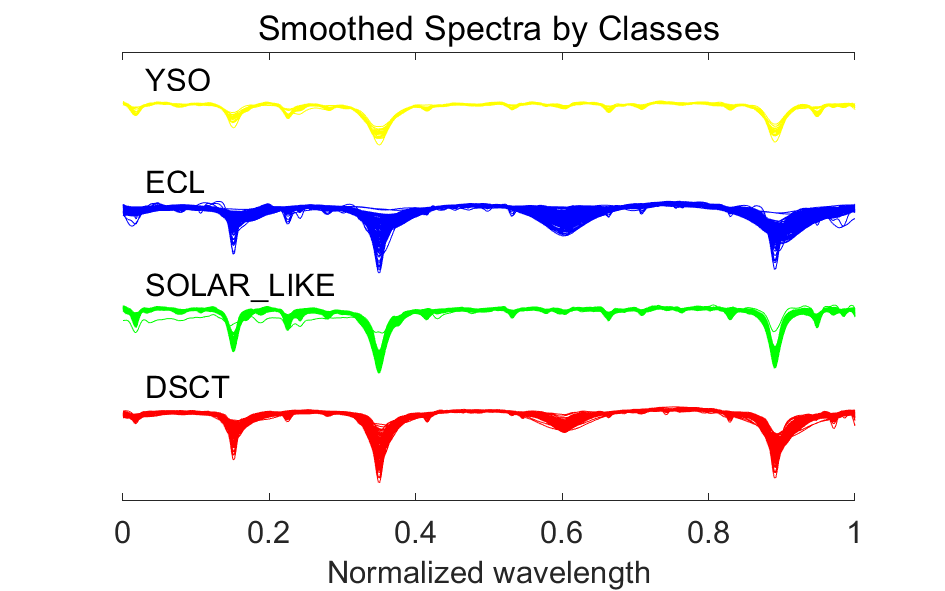}
\caption{Left: The boxplot of the best class scores of each class; Right: The smoothed spectra of each class on the window normalized to $[0,1]$. The flux values are vertically shifted according to different classes. }\label{fg_real_score}
\end{figure}

As we mentioned above, the class of each spectrum is not determined by experts but a score produced by a machine learning pipeline \citep{Rimoldini2023}. The box plot of the best class scores of the interested classes is given in the left panel of Figure~\ref{fg_real_score}. We see that although DSCT accounts for 36\% of all spectra, the average score of DSCT is the lowest. We regard the spectra with the score larger than 0.6 as labeled individuals and the rest as unlabeled individuals. Although the labels are not genuinely true, they provide a reasonable sample for validation. This leads to the labeled sample of size $n_{\ell,\textrm{fix}}=575$, including 173 DSCT ($Y=0$), 231 SOLAR\_LIKE ($Y=1$), 150 ECL ($Y=2$), and 21 YSO ($Y=3$), and an unlabeled sample of $n_{u,\textrm{fix}}=2406$ spectra. The (vertically shifted) smoothed spectra of different classes in the labeled sample are plotted in the right panel of Figure~\ref{fg_real_score}.

To compute the Fermat distance, we set $\alpha=2$ and utilize union of $200$-NN graph and the MST graph as the adjacency graph. The value of $k$ is set as $[n_\ell/10]$ for all $k$-NN related classifiers, where $n_\ell$ is specified below. To implement SVM with the Fermat distance and multiple classes, we utilize the ``one-vs-all'' strategy \citep{Hastie2009}. The weights $1/\pi_k$ are assigned to individuals with $Y=k$ to mitigate the effect of imbalanced classes, where $\pi_k$ is defined below. The estimated intrinsic dimension $\hat{d}=4$ is obtained by the approach in \citet{Tan2024}.

\begin{figure}[t]
\centering
\includegraphics[width=0.6\textwidth]{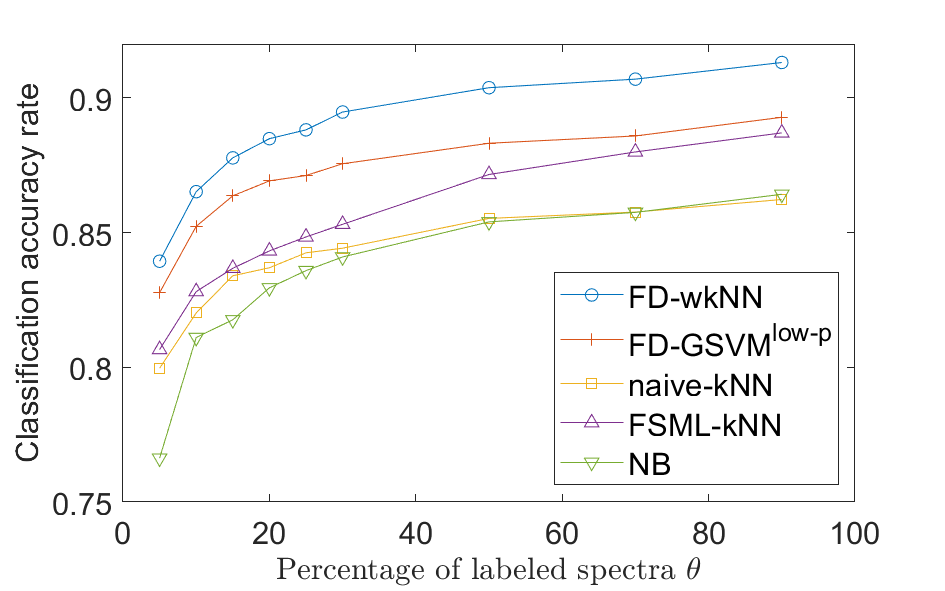} 
\caption{Average rate of classification accuracy of all classifiers applied on the astronomical spectral data under different values of $\theta$.}\label{fg_real_acc}
\end{figure}

We still consider all the classifiers as in Section~\ref{sc_sim_1}. To compare classification accuracy among different approaches, we further randomly select $[\pi_y  n_{\ell,\textrm{fix}}\times\theta\%]$ out of individuals in class $Y=y$, for $y=0,1,2$ and $3$, where $\pi_y$ is the empirical prior probability of class $y$ and $\theta=5, 10, 15, 20, 25, 30, 50, 70 $ and $90$. Specifically, the class DSCT ($Y=0$) corresponds to $\pi_0=173/575$, and the other $\pi_y$'s are defined similarly. The training sample thus contains $n_\ell = \sum_{y=0}^3[\pi_y n_{\ell,\textrm{fix}}\times\theta\%]$ individuals, and the rest $n_{\ell,\textrm{fix}}-n_{\ell}$ individuals in the labeled sample are used as the testing sample. For each of the classifiers, the classification accuracy is based on comparing its estimates and high-confidence proxy labels (i.e., those with scores larger than 0.6) from $n_{\ell,\textrm{fix}}-n_{\ell}$ individuals in the testing sample. When training our semi-supervised classifiers, all the $n_u=n_{\ell,\textrm{fix}}-n_{\ell}+n_{u,\textrm{fix}}$ spectra are included as the unlabeled sample, whereas the unlabeled sample cannot be used in training the supervised classifiers. The procedure is repeated 100 times for each $\theta$, and the averaged classification accuracy of all classifiers is shown in Figure~\ref{fg_real_acc}.

As illustrated in Figure~\ref{fg_real_acc}, both our semi-supervised classifiers outperform all supervised classifiers under all possible $\theta$. In particular, FD-w$k$NN performs the best, followed by FD-GSVM$^{\textrm{low}-p}$. Compared to NB that performs the worst under this dataset, the classification accuracy of FD-w$k$NN is improved by roughly $5\%$ to $7\%$, which shows promising usefulness of the Fermat distance exploiting the unlabeled sample. Among the supervised classifiers, FSML-$k$NN, which is designed for manifold-support functional data, performs the best. This strengthens the plausibility that the spectra admit a low-dimensional manifold structure.

\begin{figure}[t]
\centering
\includegraphics[width=0.7\textwidth]{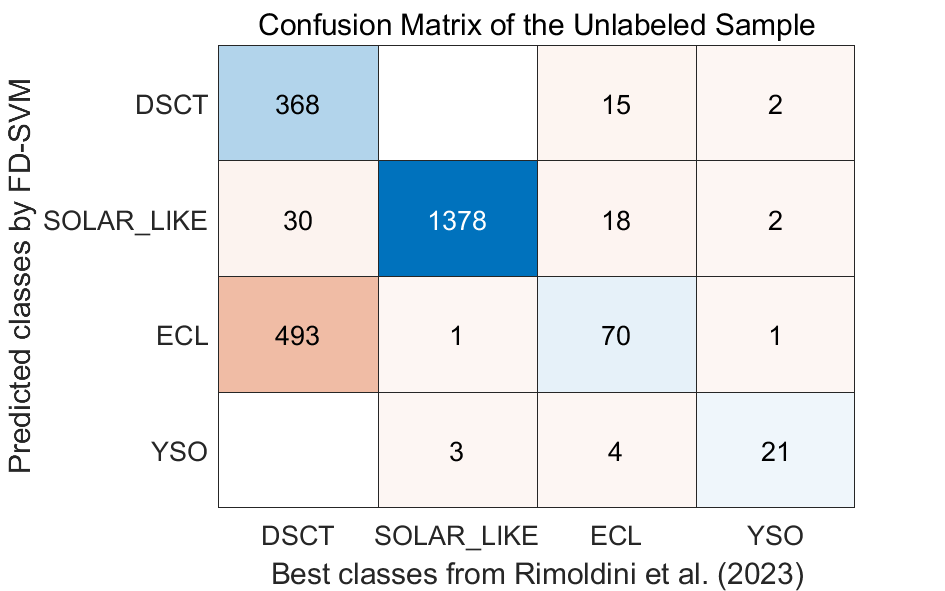} 
\caption{Confusion matrix of the classes predicted by FD-w$k$NN  and the corresponding best classes from~\citet{Rimoldini2023} for the spectra with the score smaller than 0.6.}\label{fg_real_confs}
\end{figure}

Finally, we compare the FD-w$k$NN predictions for the unlabeled sample with the corresponding best classes from \citet{Rimoldini2023}. We select FD-w$k$NN because it performs best in the evaluation above. We train it using all $n_{\ell,\rm{fix}}+n_{u,\rm{fix}}=2981$ spectra, of which $n_{\ell,\rm{fix}}=575$ spectra are labeled, and predict the classes of the remaining $n_{u,\rm{fix}}$ spectra. The tuning parameters, including $k$, $k_g$ and $\alpha$, are the same as above. Figure~\ref{fg_real_confs} shows the confusion matrix comparing the predicted and best classes in~\citet{Rimoldini2023}.  

We see from Figure~\ref{fg_real_confs} the main discrepancy is that FD-w$k$NN assigns the class ECL to 493 spectra which are labeled as DSCT in \citet{Rimoldini2023}. This is not surprising due to several reasons: (1) As shown in the left panel of Figure~\ref{fg_real_score}, the average score of DSCT is the lowest, indicating that \citet{Rimoldini2023}'s approach labels those spectra as DSCT with low confidence. (2) As mentioned in Section~\ref{sc_real_des}, it is well known by astronomers that the absorption line profiles of DSCT and ECL often look similar, making them notoriously difficult to be distinguished. (3) Moreover, the distributed classifier in \citet{Rimoldini2023} is trained on the photometric time-series data, which are distinct from the spectra used here. Given these reasons and the strong consistency of the confusion matrix excluding the $(3,1)$ entry, we conclude that the results produced by FD-w$k$NN are actually satisfactory.

\section{Discussion}\label{sc_dis}

In this work, we propose a density-sensitive semi-supervised classification framework for discrete and noisy functional data. By extending the Fermat distance to the functional domain, we develop distance-based semi-supervised classifiers that explicitly exploit the manifold and cluster assumptions. Our estimation procedure is designed for computational efficiency to handle massive data commonly seen in semi-supervised applications. In theory, we derive convergence rates of the expected excess risk within clusters of the weighted $k$-NN classifier under both true and estimated Fermat distances. The main theoretical contribution is to propagate
reconstruction error through the graph-based estimator and the resulting within-cluster
excess-risk bound. The resulting upper bound separates the manifold approximation and trajectory reconstruction terms and identifies conditions under which the latter is asymptotically negligible. For both simulated and real spectral data, we demonstrate that our semi-supervised classifiers yield more accurate classification compared with existing supervised classifiers.

Our semi-supervised classifiers are distance-based. The distance estimation is not improved by the unlabeled sample, if the estimation error from the functional trajectory reconstruction dominates. This leads to specific rate requirements for the unlabeled sample to be potentially beneficial in our methodology. On the other hand, it is natural to ask whether it is possible to develop semi-supervised approaches where the unlabeled sample is beneficial under sparse or moderate designs, which motivates future research. Moreover, considering the functional semi-supervised framework for more complex tasks such as regression and statistical inference represents exciting topics for future exploration.

\linespread{1.5}\selectfont
\bibliography{FSML_ref}

\end{document}